# A Knowledge graph representation of baseline characteristics for the Dutch proton therapy research registry.

Authors: Matthijs Sloep* [1] [0000-0003-3602-1885], Petros Kalendralis* [1] [0000-0002-4471-2021], Ananya Choudhury[1] [0000-0002-0422-7996], Lerau Seyben[1] [0000-0003-2465-8512], Jasper Snel[1] [0000-0001-9518-7274], Nibin Moni George[1] [0000-0002-0593-8693], Martijn Veening[2] [0000-0002-1118-6977], Johannes A. Langendijk[2] [0000-0003-1083-372X], Andre Dekker[1,3] [0000-0002-0422-7996], Johan van Soest[1,3] [0000-0003-2548-0330], Rianne Fijten[1] [0000-0002-1964-6317]

1. Department of Radiation Oncology (Maastro), GROW School for Oncology, Maastricht University Medical Centre+, Maastricht, The Netherlands
2. Department of Radiation Oncology, University of Groningen, University Medical Centre Groningen. Groningen. The Netherlands.
3. Brightlands Institute of Smart Society (BISS), Faculty of Science and Engineering, Maastricht University, Heerlen, The Netherlands

*Equally contributed

## Abstract

Cancer registries collect multisource data and provide valuable information that can lead to unique research opportunities. In the Netherlands, a registry and model-based approach (MBA) are used for the selection of patients that are eligible for proton therapy. We collected baseline characteristics including demographic, clinical, tumour and treatment information. These data were transformed into a machine readable format using the FAIR (Findable, Accessible, Interoperable, Reusable) data principles and resulted in a knowledge graph with baseline characteristics of proton therapy patients. With this approach, we enable the possibility of linking external data sources and optimal flexibility to easily adapt the data structure of the existing knowledge graph to the needs of the clinic.

## Introduction

Proton therapy has emerged as a novel treatment modality that has the potential to reduce toxicity rates and further improve tumour control due to the depth-dose characteristics of the proton particle[1]. Because of scarcity and costs, the Netherlands has initiated the development of a model-based approach (MBA) to select those patients for proton therapy that will benefit the most[2]. By comparing the dose difference in the organs at risk (OARs) in delta Normal Tissue Complication Probability (ΔNTCP) models for photon and proton plans and the resulting 3D radiation dose, the MBA estimates the potential benefit for an individual patient.

To ensure the MBA remains valid, one needs to continuously update and validate these ΔNTCP models. In the Netherlands, the ProTRAIT (**PRO**ton **T**herapy **R**ese**A**rch reg**IsT**ry) initiative has set up a national registry that collects real world data from patients previously treated with proton or photon radiotherapy. The initiative's aim is to systematically and automatically register these data and its ultimate goals are to minimise radiation-induced toxicities in the healthy tissues, to improve quality of life, and to escalate the dose to target (tumor) cells.

In order to fulfill the ProTRAIT initiative's aim, it is important to develop an architecture that can handle the semi-structured nature of radiotherapy data and adheres to the FAIR (Findable, Accessible, Interoperable, Reusable) principles. The authors who first published the principles point out the need to improve infrastructures to support the reuse of (scholarly) data and created guidelines to facilitate this[3]. Taking into account the semi-structured and multisource nature of radiotherapy data (imaging, biological and clinical data), as well as the heterogeneous clinical workflows and data analysis pipelines between individual centres, the implementation of the FAIR principles can enable a standardised framework for data management and processing. Furthermore, the accessibility, interoperability and reusability aspect of FAIR will enable a quicker external and independent validation of research findings[3].

FAIR data are often collected in a semantic data model and represented in a knowledge graph[4]. Literature defines a (biomedical) knowledge graph as "a resource that integrates one or more expert-derived sources of information into a graph where nodes represent biomedical entities and edges represent relationships between those entities"[5]. In biomedical science, knowledge graphs are often built based upon an existing database. In our case, the ProTRAIT data registry graph had to be built manually, however inspired by previous work on the Radiation Oncology Ontology[6].

Up until this moment the data needed for the ProTRAIT registry are manually collected in each individual centre and then transferred to a centralised electronic data capture (EDC) system. A vital part of the upload infrastructure is choosing an interoperable data model; this means choosing a data structure. In this paper we present our choice for a domain specific knowledge graph that stores relevant observational patient data into an interoperable, machine readable format. The knowledge graph specification is publicly available (DOI:10.5281/zenodo.5060069[7]) but the data itself is not due to privacy and legal requirements associated with patient data; access to these data will be formalized in the near future by the ProTRAIT consortium.

## Methods and Results

The clinical items listed in the registry and graph were selected by domain experts based on established clinical workflows and were subsequently reviewed by the relevant national expert community: part of the "Nederlandse Vereniging voor Radiotherapie en Oncologie." - the Dutch national association for radiotherapy and oncology. This resulted in a flat data model, a list of items with little to no relation between the individual elements. The structure of this flat data model will be discussed in depth in the next paragraphs. The items in this list and others can be found on the [Github repository](), including the definition and required data element type (integer, string, date etc.) in .xlsx format.

In our case the construction of the graph and ontology classes was a coordinated effort by medical physicists, physicians and computer scientists. Their combined expertise was used to define nodes and edges in the graph. The knowledge graph was modeled using the Resource Description Framework (RDF), a World Wide Web Consortium data standard (W3C). It is originally designed for metadata but also used for knowledge management applications[5]. The RDF format is based on the representation of the data in triples format (Subject-Predicate-Object, eg. Patient-has Disease-Neoplasm). The structure of the knowledge graph was constructed with the patient class at the centre with connections to different sections of clinical information, such as the age and biological sex, and information regarding the baseline treatment (eg. date of first radiotherapy course).

The list of clinical items were represented in the R2RML mapping language[8] to describe the graph structure, and to facilitate the data conversion process. In this graph structure, several publicly available ontologies related to the radiation oncology field, bundled in the Radiation Oncology Ontology (ROO)[6], were reused. The importance of ontologies specific to the radiotherapy domain has been highlighted by Phillips et al.[9]. The use of ontologies enhances the data interoperability and reusability with a clear definition of the different data classes including knowledge representation.

Figure 1 shows the baseline characteristics in our knowledge graph with all the nodes and edges designed by domain experts. The different data classes can be grouped in three different categories; (I) baseline characteristics and demographic information, (II) baseline tumour, and (III) radiotherapy planning information. In this visualisation of the graph, we present the relation between variables connected to a patient that all together make up our generic list. Furthermore, we would like to underline that additional data can be linked to this graph easily. The structure of the knowledge graph is open source and the R2RML mapping files can be found on http://www.protrait.nl (licence: CC-BY)

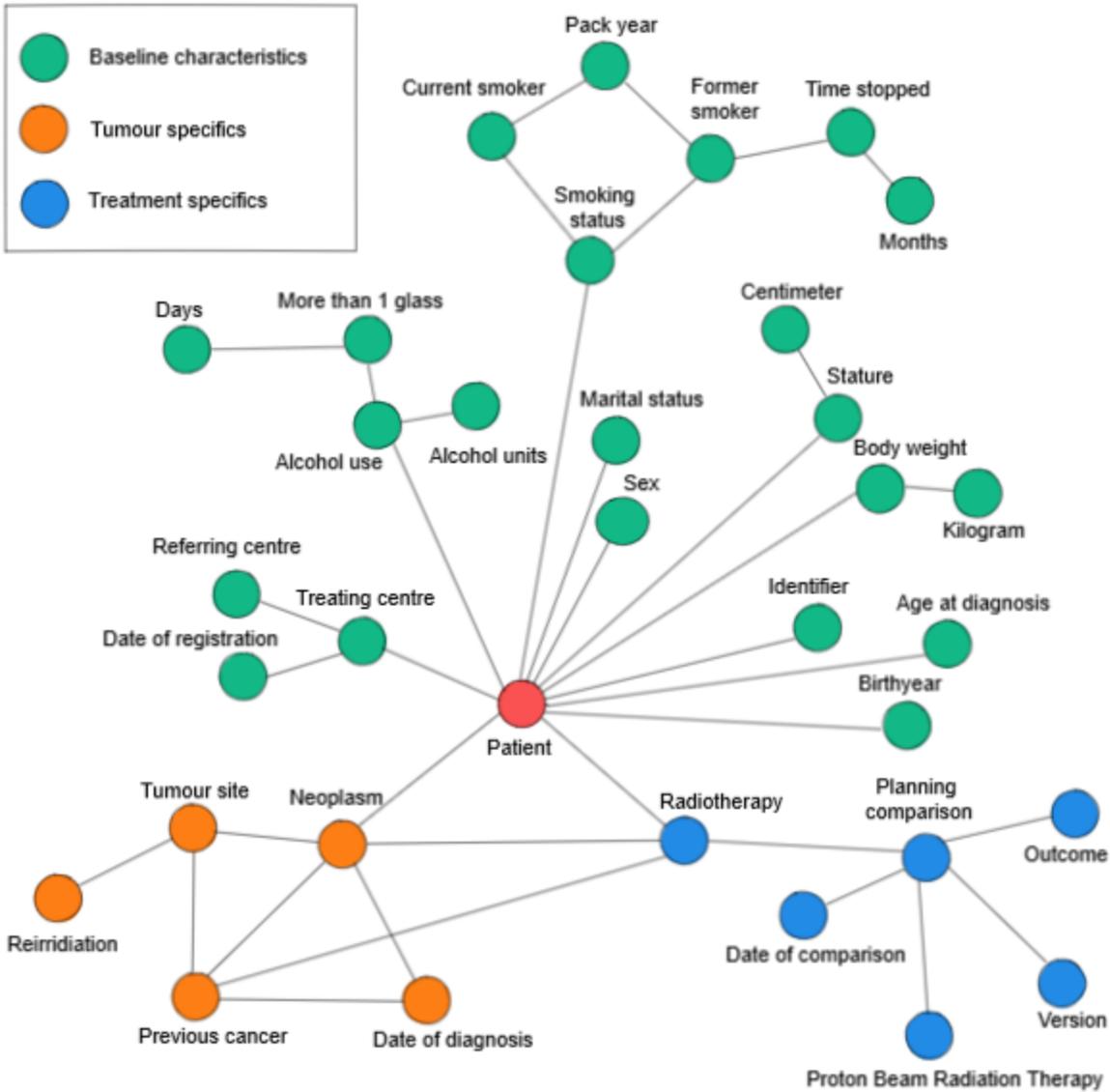

Figure 1: Visualisation of the knowledge graph created with the baseline characteristics (green) tumour specific variables (orange) and treatment variables (blue) of the patients eligible for proton radiotherapy. For readability purposes we have excluded the predicates between the different instances and classes.

## Discussion

In this study, we developed a knowledge graph to store clinical patient characteristics for the proton therapy registry. We chose this data model because of the specific characteristics of the ProTRAIT project. Our knowledge graph was based on the RDF[10] data model using the R2RML[8] mapping language and publicly available ontologies[6,9,11] as it facilitates the interoperability and reusability of data .

Proton therapy is a relatively new treatment and its indications and application are likely to change when new insights develop. Hence, the data model and structure must be designed with flexibility and a transient practical application in mind: as proton therapy gains salience in The Netherlands, new clinical applications will appear and there will be a shift in the threshold of which patients can undergo treatment because of the limited treatment capacity. New models are developed to tackle this shift and for this reason, our architecture must be able to easily adapt to new data elements and model transformations. The semantic data model is flexible because we can define and add ontology classes and their definitions are shared and accessible to others in line with the FAIR principles, which gained traction in the radiotherapy world[12], while still keeping them backward compatible. The semantic data model was developed with machine readability and thus exchange and interoperability in mind. The flexibility further shows in adding new variables to the data model and the cardinality limitations that relational databases have. There is no need to create new tables to tackle {one/many}-to-many relations; for instance, additional treatments can exist in the same graph as additional instances of the treatment class. Finally, the ease to which multiple datasets/data sources can be queried (eg. third parties datasets) in a single unified query is an additional point that makes the knowledge graph an advantageous data format over a relational database[6].

Knowledge graphs are not mainstream in clinical data capture systems. Indeed, relational databases still are widely used for clinical data storage. However, knowledge graphs have significant advantages over relational databases, such as flexibility and the ease with which semantic data may be enriched. Most important, perhaps, is that the ProTRAIT data model stands out in interoperability. As hospitals generally use local syntaxes for data registration, their relational databases are not interoperable.

An alternative to our RDF based ProTRAIT approach could have been the Observational Medical Outcomes Partnership (OMOP)[13]. OMOP is a common data model and technical architecture, which works in a similar manner to our FAIR approach initiative to collect observational data using relational databases. OMOP supports population exchange, but not with the flexibility that FAIR or RDF representations have. Semantic integration adds context, uncertainty and detail to the data annotation on a level that OMOP cannot[4]. Furthermore, OMOP is not designed for handling detailed procedural information in a structured and standardized manner. Health Level 7 (HL7) is the clinical standard that describes data formats and elements is a relevant standardisation initiative; the latest version, Fast Healthcare Interoperability Resources (FHIR), focuses on communication and information exchange[13]. FHIR is designed for electronic health record (EHR) based sharing of data from individual patients between institutions and is broadly supported[14]. However, since FHIR has limited functionality in exchanging population level data we opted for the semantic data model.

The metadata that ontologies add to the knowledge graph not only make the data adhere to the FAIR principles but also enrich the data and serve another practical purpose. By using domain specific ontologies in our knowledge graph, the original real world data can co-exist in the same graph together with the project specific categories and numerical values. In the analysis there is the potential to allow algorithms to infer indirect knowledge from the graph, which is not possible in a flat relational database. In other words, the clinical expertise in the creation of the ontology and knowledge graph means that the metadata is enhancing the instance data, and that inferencing could potentially improve AI/ML analysis of the data. For example, identification of similar patient groups, depending on their characteristics, may enable a personalised approach for prognostic studies.

In the future the central registry may become substituted by a federated/distributed analysis of data using the Personal Health Train (PHT)[15]. The requirements set by the Semantic Web technologies allows machines to understand and interpret the data element classes. Furthermore, ML applications can be implemented, such as the validation and exchange of prediction models using the privacy preserving PHT infrastructure[15]. Moreover, the knowledge graph format may efficiently serve data handling and storage of (distributed) large-scale datasets ("big data").

# Conclusion

With this study we present our knowledge graph; a database solution for a clinical and research repository that ensures a high degree of flexibility which is needed in a new and advancing field. Our research repository promotes adherence to the FAIR principles. This will facilitate re-use of the data for instance by linking the data to other data sets or incorporating the PHT infrastructure for federated learning analysis. Lastly, the knowledge graph enhances the data and creates opportunities for improved Machine Learning (ML)/Artificial Intelligence (AI) analysis. Future plans are to link sets of tumour specific items that contain data elements related to the treatment, patient reported outcome measures and radiotherapy dose information in order to allow for the design and validation of ΔNTCP models needed in the MBA proton patients' selection.

# Acknowledgements

We want to thank the Dutch Cancer Society KWF 30943405N for funding the development of the proTRAIT infrastructure and subsequent proton therapy research repository.